\documentclass[modern]{emulateapj}
\received{December 14, 2017}
\revised{January 12, 2018}
\accepted{January 16, 2018}

\shorttitle{Perseus Arm Dynamics}
\shortauthors{Baba et al.}

\begin{document}

\title{Gaia DR1 evidence of disrupting Perseus Arm}

\email{jun.baba@nao.ac.jp; d.kawata@ucl.ac.uk}

 \author{
 Junichi Baba$^1$,
 Daisuke Kawata$^2$,
 Noriyuki Matsunaga$^3$,
 Robert J. J. Grand$^{4,5}$, 
 and
 Jason A. S. Hunt$^6$
 }
 \affiliation{
 $^1$ National Astronomical Observatory of Japan, Mitaka, Tokyo 181-8588, Japan\\
 $^2$ Mullard Space Science Laboratory, University College London, Holmbury St. Mary, Dorking, Surrey RH5 6NT, UK\\
 $^3$ Department of Astronomy, The University of Tokyo, 7-3-1 Hongo, Bunkyo-ku, Tokyo 113-0033, Japan\\
 $^4$ Heidelberger Institut f\"{u}r Theoretische Studien, Schloss-Wolfsbrunnenweg 35, 69118 Heidelberg, Germany\\
 $^5$ Zentrum fur Astronomie der Universitat Heidelberg, Astronomisches Recheninstitut, Monchhofstr 12-14, D-69120 Heidelberg, Germany\\
 $^6$ Dunlap Institute for Astronomy and Astrophysics, University of Toronto, Ontario M5S 3H4, Canada\\
 }

%
%
%
%



\begin{abstract}
We have discovered a clear sign of the disruption phase of the Perseus arm in the Milky Way using Cepheid variables, taking advantage of the accurately measured distances of Cepheids and the proper motions from Gaia Data Release 1. Both the Galactocentric radial and rotation velocities of 77 Cepheids within 1.5~kpc of the Perseus arm are correlated with their distances from the locus of the Perseus arm, as the trailing side is rotating faster and moving inward compared to the leading side. We also found a negative vertex deviation for the Cepheids on the trailing side, $-27.6\pm2.4$~deg, in contrast to the positive vertex deviation in the solar neighborhood. This is, to our knowledge, the first direct evidence that the vertex deviation around the Perseus arm is affected by the spiral arm. We compared these observational trends with our $N$-body/hydrodynamics simulations based on a static density-wave spiral scenario and those based on a transient dynamic spiral scenario. Although our comparisons are limited to qualitative trends, they strongly favor a conclusion that the Perseus arm is in the disruption phase of a transient arm. 
\end{abstract}




\section{Introduction}
\label{sec:Intro}

How spiral arms in disk galaxies are created and maintained has been a long-standing question in galactic astronomy. For isolated disk galaxies there are two different theories of spiral arms which have different lifetimes \citep[][]{DobbsBaba2014}.
The quasi-stationary density wave theory (hereafter SDW arm) characterises spirals as rigidly rotating, long-lived wave patterns \citep[i.e., $\gtrsim$ 1 Gyr;][]{LinShu1964,BertinLin1996}.
On the other hand, dynamic spiral theory (hereafter DYN arm) suggests spiral arms are differentially rotating, transient, recurrent patterns on a relatively short time scale, {$\sim$}100~Myr \citep{SellwoodCarlberg1984,Baba+2009,Fujii+2011,Grand+2012a,Grand+2012b,Baba+2013, D'Onghia+2013,Baba2015}. 

Because a DYN arm is almost co-rotating with the stars at every radii, $N$-body simulation studies show that there should be a characteristic gas and stellar motion affected by the spiral arms \citep{Grand+2012a,Grand+2016b,Baba+2013}. \citet{Kawata+2014} suggested that comparing the stellar velocity properties between the trailing and leading side of a spiral arm would provide crucial information regarding its origin \citep[see also][]{Hunt+2015}. The Gaia mission \citep{GaiaMission016} recently published its first data release \citep[DR1;][]{GaiaDR1}, including the proper motions and parallaxes for two million bright stars in common with the Tycho-2 catalogue, known as the Tycho-Gaia astrometric solution \citep[TGAS;][]{Michalik+2015,Lindegren+2016}.
\citet{Hunt+2017} reached a tentative conclusion favoring a DYN arm by finding a group of stars whose Galactocentric rotation velocity is unexpectedly high owing to the torque from the Perseus arm. However, the feature was observed in stars in the solar neighbourhood (distance $<0.6$~kpc) which is still far away from the Perseus arm. Hence, it is difficult to conclude whether or not this feature is due to the Perseus arm. 

Here we investigate kinematics of Cepheid variables. They are bright variable stars and their distances are accurately measured thanks to their well-calibrated period-luminosity relation \citep{Inno+2013}. 
Cepheids are also young stars whose ages are expected to be around $20-300$~Myr \citep{Bono+2005}. 
Such young stars are expected to have small velocity dispersion, and it is easier to find a systematic motion due to dynamical effects if it exists. Moreover, the age range of Cepheids is comparable to the lifetime of the DYN arm, and thus they are expected to be sensitive to the dynamical state of the spiral arms. Hence, Cepheids are a great tracer for testing the spiral arm scenario \citep[see also][]{Fernandez+2001,Griv+2017}. Thus, this {\it Letter} uses Cepheids around the Perseus arm to study the dynamical state of the Perseus arm. 

Section~\ref{sec:ObsTrend} describes our sample of Cepheids, and shows their kinematic properties. Section~\ref{sec:CompModel} presents the results of comparisons between the observed Cepheids kinematics and what is seen in the simulations with different spiral models.

\section{Peculiar Motions of Cepheids}
\label{sec:ObsTrend}


We selected a sample of Cepheids in \citet{Genovali+2014} where the distances were determined homogeneously by using near-infrared photometric data sets \citep[also see][]{Inno+2013}. Errors in distance modulus are estimated by \citet{Genovali+2014} to be 0.05--0.07~mag for most of the Cepheids. We then cross-matched this sample with the TGAS catalog and with a sample of Cepheids whose radial velocity are provided in \citet{Melnik+2015} using TOPCAT \citep{Taylor2005}, giving a collection of 206 Cepheids with known locations and kinematics. We further limit the sample based on vertical position with respective to the Sun, $|z_{\rm e,max}|<0.5$ kpc, where to take into account the error we define $z_{\rm e,max}=\sin({b})10^{(\mu_{\rm dm}+\mu_{\rm dm,e}+5.0)/5.0}/1000$~kpc, where $b$, $\mu_{\rm dm}$, and $\mu_{\rm dm,e}$ are the Galactic latitude, the distance modulus and its error in magnitude, respectively. This limit was applied to eliminate clear outliers, although our sample shows a clear concentration around the Galactic plane, with more than 70~\% being located within 100~pc, as expected for young stars like Cepheids. 

To eliminate the data with a large velocity or distance uncertainty, we discard the data whose uncertainty in velocity, $\sigma_V=\sqrt{\sigma_{V_{\rm lon}}^2+\sigma_{V_{\rm lat}}^2+\sigma_{V_{\rm HRV}}^2}$, is larger than 20~km~s$^{-1}$, where $\sigma_{V_{\rm lon}}$, $\sigma_{V_{\rm lat}}$ and $\sigma_{V_{\rm HRV}}$ are the uncertainties of the velocity measurements in the direction of longitude, $V_{\rm lon}$, latitude, $V_{\rm lat}$, and heliocentric radial velocity, $V_{\rm HRV}$. $\sigma_{V_{\rm lon}}$ and $\sigma_{V_{\rm lat}}$ are computed by taking the standard deviation of the Monte-Carlo (MC) sampling of $V_{\rm lon}$ and $V_{\rm lat}$, computed for randomly selected right ascension (RA) and declination (DEC) proper motions, using the 2D Gaussian probability distribution with their measured mean, standard error and correlation between the RA and DEC proper motions, and distance from the Gaussian probability distribution of distance modulus with the mean of $\mu_{\rm dm}$ and a standard deviation of $\mu_{\rm dm,e}$. These selections left 191 Cepheids in our sample. 

\begin{figure}
\begin{center}
\includegraphics[width=0.45\textwidth]{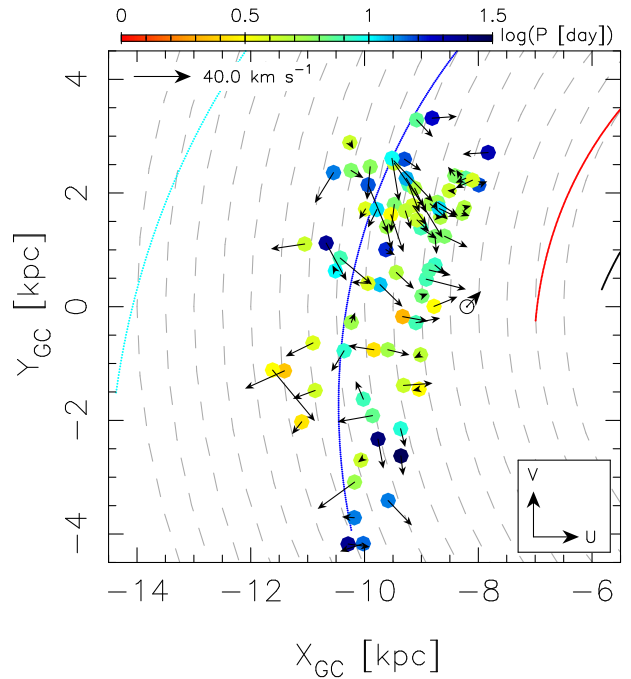}
\end{center}
\caption{Face-on distribution of the selected Cepheids 
with arrows describing their peculiar velocities with color indicating their pulsation period ($\log P$).
The cyan, blue, red and black solid lines show the positions of 
the Outer, Perseus, Sagittarius and Scutum spiral arms, respectively, measured in \citet{Reid+2014a}. 
The open circle indicates the position of the Sun and the arrow shows its peculiar motion. 
\label{fig:ObsCephMap}}
\end{figure}

From this sample, we choose Cepheids around the Perseus arm. We adopt the position of the Perseus arm locus as determined by \citet{Reid+2014a}, which provides the distance and angle to the reference point from the Sun and the pitch angle of the locus of the arm. We computed the mean distance, but projected  on the Galactic plane, between the closest point of the locus of the Perseus arm and the Cepheids in our sample, $d_{\rm Per}$, by MC sampling of the distance modulus for Cepheids and the distance between the Sun and the Galactic center of $R_0=8.2\pm0.1$~kpc \citep{Bland-HawthornGerhard2016}\footnote{In our MC sampling, for simplicity we fixed the pitch angle, but only changed the Galactocentric radius of the reference point of the Perseus arm for a sampled $R_0$, although the pitch angle also depends on $R_0$.}. We selected Cepheids within $|d_{\rm Per}|<1.5$~kpc, which results in a final catalogue of 77 Cepheids (see Figure~\ref{fig:ObsCephMap}).

We compute the Galactocentric radial velocity, $U_{\rm pec}$, and rotation velocity, $V_{\rm pec}$, after subtracting the circular velocity of the disk at the location of each Cepheid.  Again, we used 10,000 MC samples to estimate the uncertainties of $U_{\rm pec}$, $V_{\rm pec}$ and $d_{\rm Per}$, taking into account the mean and uncertainties of the distance modulus and proper motion for individual Cepheids and all the relevant Galactic parameters, such as $R_0=8.2\pm0.1$~kpc, the angular velocity of the Sun with respect to the Galactic center, $\Omega_{\rm \sun}=30.24\pm0.12$~km~s$^{-1}$~kpc$^{-1}$, the solar peculiar motion with respect to the Local Standard of the Rest, ($U_{\sun}, V_{\sun}, W_{\sun}) = (10.0\pm1.0, 11.0\pm2.0, 7.0\pm0.5$)~km~s$^{-1}$ \citep{Bland-HawthornGerhard2016} and the radial gradient of circular velocity, $dV_{\rm c}/dR=-2.4\pm1.2$~km~s$^{-1}$~kpc$^{-1}$ \citep{FeastWhitelock1997}. We take the mean and standard deviation of $U_{\rm pec}$, $V_{\rm pec}$ and $d_{\rm Per}$ from the MC sample.

\begin{figure*}
\begin{center}
\includegraphics[width=0.45\textwidth]{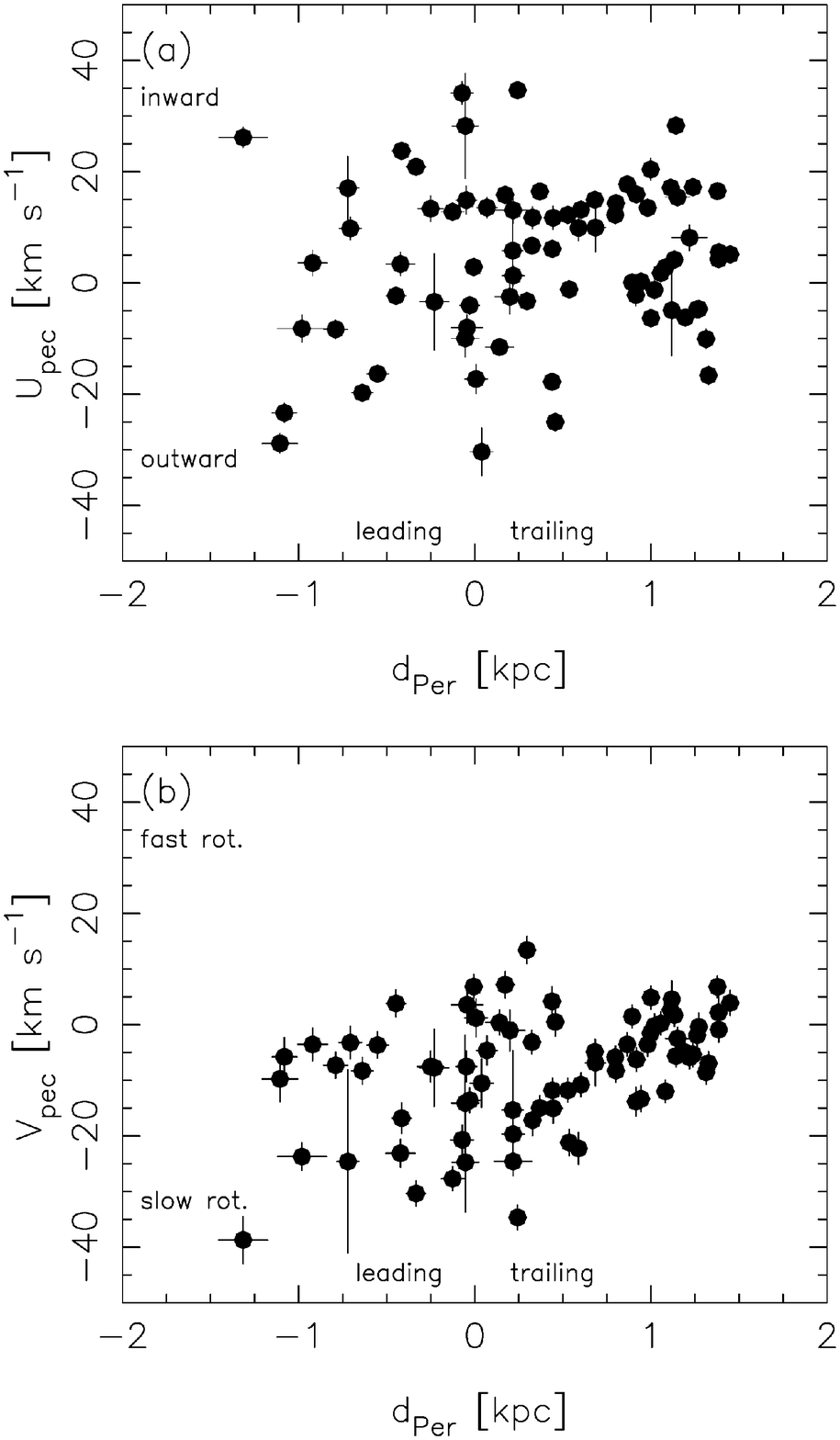}
\includegraphics[width=0.45\textwidth]{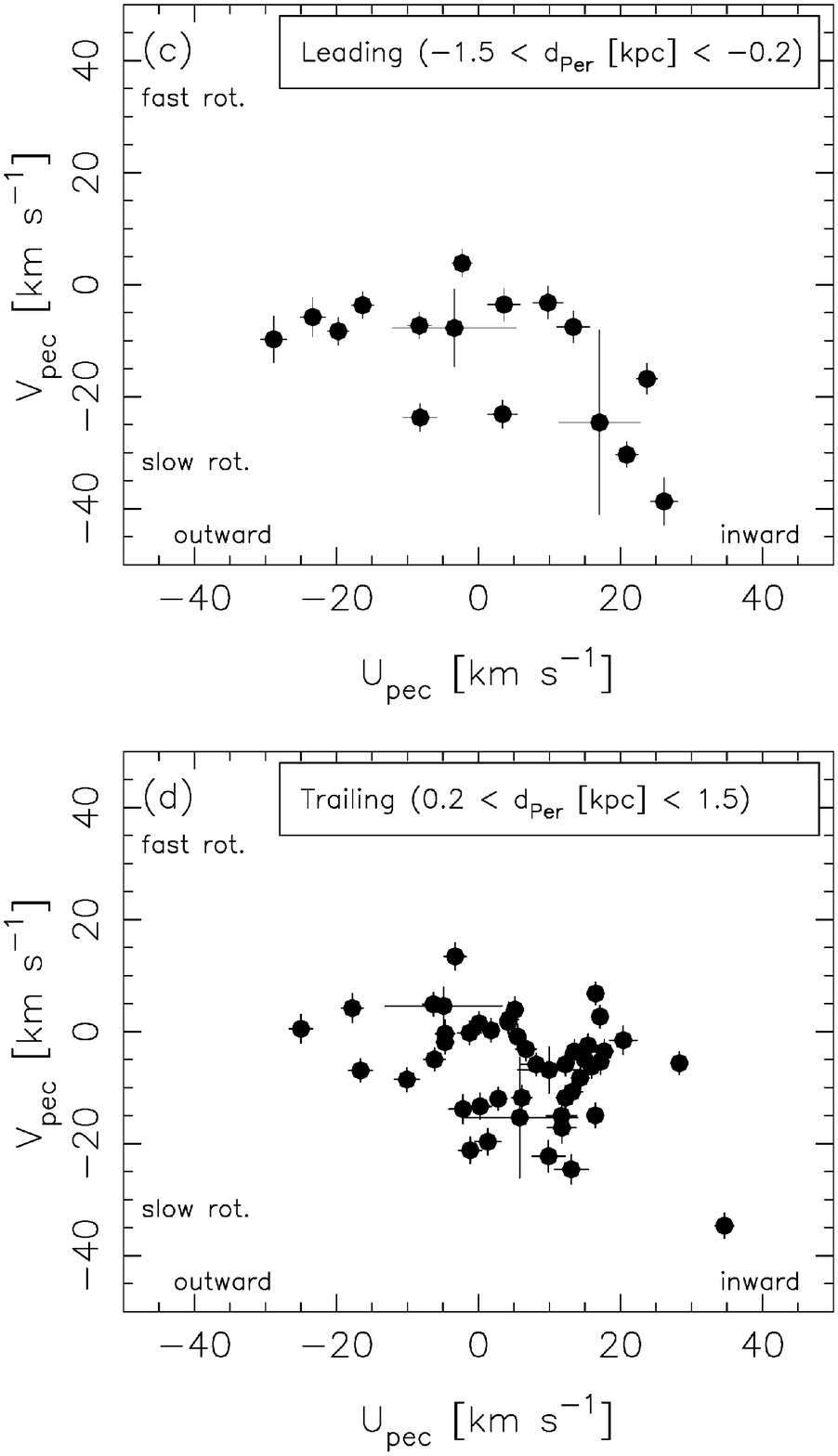}
\end{center}
\caption{
(a) $U_{\rm pec}$-$d_{\rm Pec}$ and (b) $V_{\rm pec}$-$d_{\rm Per}$ distributions 
of our Cepheid sample.
Note that $U_{\rm pec}$ is positive in the direction toward the Galactic center, 
and $V_{\rm pec}$ is positive in the direction of the Galactic rotation. 
(c)(d) $U_{\rm pec}$-$V_{\rm pec}$ distributions of our Cepheid sample 
in leading and trailing sides of the Perseus arm, respectively.
\label{fig:ObsCephStat}
}
\end{figure*}

\begin{deluxetable*}{cccccc}
\tablecaption{
Kinematics of Cepheids and Model results \label{tab:ObsModres}
}
\tablehead{
\colhead{} 
	& \colhead{Cepheids (number)}
 	& \colhead{SDW ($R_{\rm cr}=8$~kpc)} 
     & \colhead{SDW ($R_{\rm cr}=16$~kpc)}
     & \colhead{DYN ($t = 2.59$~Gyr)}  
     & \colhead{DYN ($t = 2.62$~Gyr)}
}
\startdata
$U_{\rm pec}-d_{\rm Per}$ Corr. 
	& $0.14\pm0.02$ (77) 
    & $0.21\pm0.07$
    & $-0.80\pm0.02$
    & $-0.21\pm0.03$
    & $0.14\pm0.03$\\
$V_{\rm pec}-d_{\rm Per}$ Corr. 
	& $0.40\pm0.03$ (77) 
    & $-0.34\pm0.03$
    & $-0.47\pm0.03$
    & $0.062\pm0.04$
    & $0.15\pm0.03$\\
\multicolumn{6}{c}{Trailing side ($0.2<d_{\rm Per}<1.5$ kpc)}\\
$\langle U_{\rm pec}\rangle$ (km~s$^{-1}$) 
	& $6.1\pm1.0$ (47) 
    & $-7.0\pm0.5$
    & $-24.3\pm1.8$
    & $-11.4\pm0.9$
    & $4.6\pm0.8$\\
$\langle V_{\rm pec}\rangle$ (km~s$^{-1}$)
	& $-6.3\pm2.0$ (47) 
    & $-8.5\pm0.4$
    & $-8.5\pm0.7$
    & $-7.5\pm0.8$
    & $-3.8\pm0.8$\\
$l_v$ (deg) 
	& $-27.6\pm2.4$ (47)
    & $-11.5\pm5.6$
    & $3.1\pm5.1$ 
    & $27.4\pm5.8$
    & $-42.4\pm5.2$\\
\multicolumn{6}{c}{Leading side ($-1.5<d_{\rm Per}<-0.2$ kpc)}\\
$\langle U_{\rm pec}\rangle$ (km~s$^{-1}$) 
	& $0.49\pm1.2$ (16) 
    & $-10.4\pm2.9$
    & $21.5\pm0.7$
    & $-1.1\pm1.5$
    & $-3.0\pm1.4$\\
$\langle V_{\rm pec}\rangle$ (km~s$^{-1}$) 
	& $-13.4\pm2.0$ (16) 
    & $-1.7\pm1.1$
    & $0.46\pm0.54$
    & $-11.2\pm1.0$
    & $-11.9\pm0.9$\\
$l_v$ (deg) 
	& $-28.2\pm6.5$ (16)
	& $9.0\pm3.3$
    & $-1.2\pm10.9$
    & $20.8\pm2.9$
    & $22.5\pm1.9$ \\
\enddata
\end{deluxetable*}

We first look at the correlation coefficients between $U_{\rm pec}$ and $d_{\rm Per}$ and between $V_{\rm pec}$ and $d_{\rm Per}$. 
As shown in Figures~\ref{fig:ObsCephStat}a,b,
there is a significant positive correlation for both $U_{\rm pec}$ and $V_{\rm pec}$ against $d_{\rm Per}$. 
We measure the correlations for each MC sampling described above, and take the mean and dispersion of the measurements. 
Table~\ref{tab:ObsModres} shows the correlation is statistically significant even after taking into account the observational errors and uncertainties of the Galactic parameters. The correlation is stronger in $V_{\rm pec}$ than $U_{\rm pec}$. 

Following the idea of \citet{Kawata+2014}, we compared the velocity distribution of 47 Cepheids on the trailing side (defined as $0.2<d_{\rm Per}<1.5$~kpc) and that of 16 Cepheids on the leading side ($-1.5<d_{\rm Per}<-0.2$~kpc) of the Perseus arm\footnote{We excluded Cepheids within $|d_{\rm Per}|<0.2$~kpc considering the uncertainty of the locus of the Perseus arm.}. We found a significant offset in the mean velocity of these samples 
as expected from the correlation with $d_{\rm Per}$ (see also Table~\ref{tab:ObsModres}).
The mean velocity in both $\langle U_{\rm pec}\rangle$ and $\langle V_{\rm pec}\rangle$ is higher on the trailing side. To our knowledge, these results are the first statistically significant observational evidence of the difference in dynamical properties of stars on different sides of the spiral arm.

Furthermore, we calculated the vertex deviation, $l_v=0.5\times \arctan(2 \sigma_{UV}^2/(\sigma_U^2-\sigma_V^2))$ \citep[including the correction term suggested by][for the case of $\sigma_V>\sigma_U$]{VorobyovTheis2008}, where $\sigma_{UV}^2$ is the covariance between $U_{\rm pec}$ and $V_{\rm pec}$, for the sample on the trailing and leading sides (see their $U_{\rm pec}$-$V_{\rm pec}$ distribution in Figures~\ref{fig:ObsCephStat}c,d). 
The results are summarized in Table~\ref{tab:ObsModres}. On the leading side, the number of Cepheids in the sample is too small to measure $l_v$ confidently. On the trailing side, in contrast, the vertex deviation is clearly negative, which is opposite to the positive one (about $+20$ deg) of the young stars in the local solar neighborhood \citep[e.g.][]{DehnenBinney1998,Rocha-Pinto+2004}.
To our knowledge, this is the first detection of the change of sign of the vertex deviation near the spiral arms. 
Such a change is expected in various spiral arm models \citep[][]{Roca-Fabrega+2014}.

Finally, we discuss the effect of interstellar reddening and extinction considering its importance for the objects in the disk \citep{Matsunaga2017}. \citet{Inno+2013} assumed the total-to-selective reddening ratio of \citet{Cardelli+1989}, which is different from some of the recent values \citep[e.g.][]{Nishiyama+2006,Alonso-Garcia+2017}. The reddenings of the Cepheids around the Perseus arm are, however, relatively small, $E_{J-K_{\rm s}}\leq 0.5$~mag, and the uncertainty caused by the extinction law (up to 7~\% in distance) does not change our results.

\section{Dynamical Nature of the Perseus Arm}
\label{sec:CompModel}

We compare our findings with $N$-body/hydrodynamic simulations with different spiral models.
The simulations include self-gravity, radiative cooling, 
star formation, and stellar feedback \citep{Saitoh+2008,SaitohMakino2009}.
The DYN arm model is a barred spiral galaxy formed from an initial axisymmetric model, spontaneously. 
The bar is an almost stable pattern, but the amplitudes, pitch angles, 
and rotational frequencies of the spiral arms 
change within a few hundred million years \citep[][]{Baba2015}.
The SDW models are from \citet{Baba+2016} and have a rigidly rotating two-armed spiral 
(external potential) with a pitch angle of $12$~deg and a spiral amplitude of $3$~\%. 
To study the impact of the location of the co-rotation radius ($R_{\rm cr}$), 
we used two SDW models with $R_{\rm cr} = 8$~kpc \citep[e.g.][]{Fernandez+2001} and 
16~kpc \citep[e.g.][]{Lin+1969}.

\begin{figure}
\begin{center}
\includegraphics[width=0.33\textwidth]{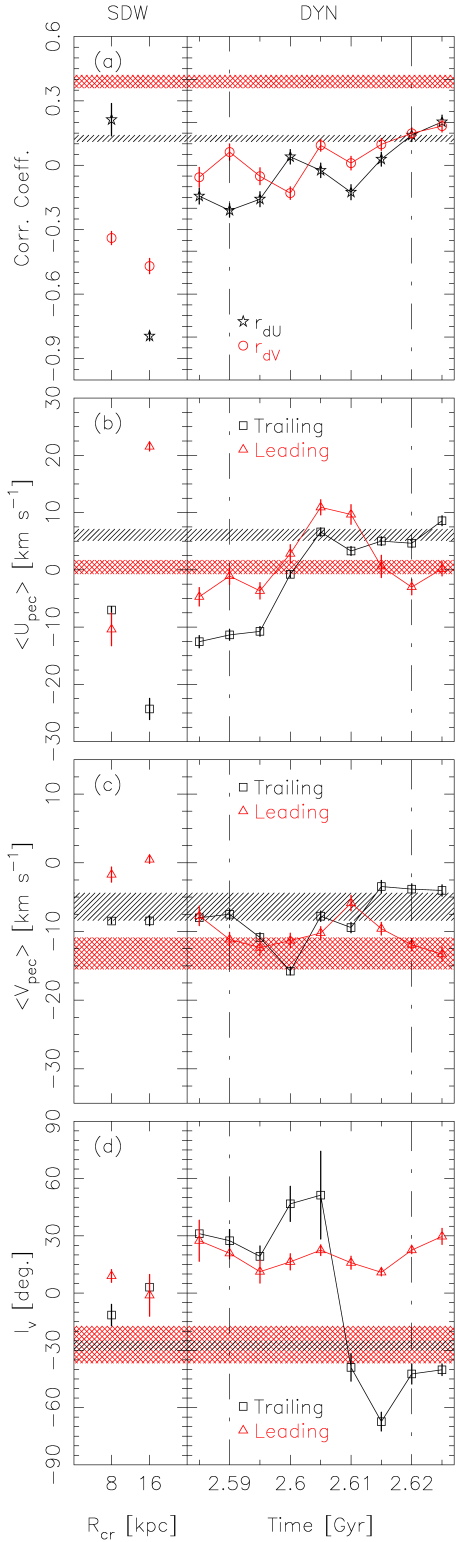}
\end{center}
\caption{
Comparison between the models and the observed kinematics of Cepheids in terms of 
correlations of $U_{\rm pec}-d_{\rm Per}$ and $V_{\rm pec}-d_{\rm Per}$ (panel a), mean $U_{\rm pec}$ (panel b), mean $V_{\rm pec}$ (panel c) and the vertex deviation, $l_v$ (panel d). Horizontal shaded areas in black and red indicate the observed values and the $1 \sigma$ uncertainty ranges for $U_{\rm pec}-d_{\rm Per}$ and $V_{\rm pec}-d_{\rm Per}$) correlation, respectively, in the top panel, while the measured values in the trailing (leading) side of the arm are indicated by different colors in the other panels. The model results are shown with the open symbols with error bars. Black stars (red square) in the top panel shows the $U_{\rm pec}-d_{\rm Per}$ ($V_{\rm pec}-d_{\rm Per}$) correlation. In the second, third and bottom panels, black square (red triangle) show the measured values in the trailing (leading) side of the arm. Note that the left side of the panels shows the SDW model results for two different $R_{\rm cr}$ values, 
whereas in the right side the DYN model results are presented as a function of time.
These results are also summarized in Table \ref{tab:ObsModres}.
\label{fig:CompModels}
}
\end{figure}

To compare the simulations with the observational data,  we have applied the same analysis as Section~\ref{sec:ObsTrend} for the simulations. First we identify a spiral arm similar to the Perseus arm in terms of the Galactocentric radial range. Then, we selected young star particles ($50$--$200$~Myr) around the arm which are located in a radial and azimuthal range similar to that of our Cepheids sample. Note that the pitch angle of our SDW models are not tuned to match the Perseus arm, but the one to best explain both the Scutum and Perseus arms with a single pitch angle. Also, the pitch angle of the DYN model is changing as time goes on. Thus, for both SDW and DYN models we measure the distance of the particles from the gas arm \citep[to be consistent with the identification of the arm by the star forming regions in][]{Reid+2014a}, $d_{\rm arm}$, irrespective of the pitch angles of the spiral arms, and consider it same as $d_{\rm Per}$ for our Cepheids data analysis. The results of the simulations are summarized in Table \ref{tab:ObsModres}, and are shown in Figure~\ref{fig:CompModels}.

We first compare the observation with the SDW models. As shown in left side panels of Figure~\ref{fig:CompModels}, SDW($R_{\rm cr}=16$) reproduces none of the observed trend, suggesting that this model is clearly rejected. On the other hand, SDW($R_{\rm cr}=8$) shows some degree of success in the positive correlation coefficient of $U_{\rm pec}$-$d_{\rm Per}$ (Figure~\ref{fig:CompModels}a) and a negative vertex deviation in the trailing side (Figure~\ref{fig:CompModels}d). However, this model fails to reproduce the positive correlation in $V_{\rm pec}$-$d_{\rm Per}$ (Figure~\ref{fig:CompModels}a). Hence, we conclude that irrespective of $R_{\rm cr}$ (i.e., the pattern speed), it is difficult for the SDW models to explain the observed features in Section~\ref{sec:ObsTrend}. This does not mean that we can reject the SDW scenario. Indeed, we have not explored models of different pitch angles and/or different strength of the arms; moreover, our SDW models do not include the bar, which is likely to affect the dynamical features \citep[e.g.][]{Monari+2016b}. Although these different kinds of models need to be tested against our observed Cepheid kinematics, the SDW spiral models tend to show a regular trend in stellar kinematics around the spiral arms \citep[e.g.][]{Roca-Fabrega+2014,Pasetto+2016,Antoja+2016}. We therefore expect that the positive correlation of $V_{\rm pec}$-$d_{\rm Per}$ is difficult to obtain in the SDW model alone. 

We then compare the DYN model with the observations. The right side panels of Figure~\ref{fig:CompModels} show the results around a spiral arm which grew and was disrupted around $t=2.59$ and $t=2.62$ Gyr (indicated with vertical dot-dashed lines), respectively. As shown in \citet{Baba+2013} and \citet{Grand+2014}, the kinematic properties of the stars around the DYN spiral arms change with time. As a result, the growing phase of the DYN arm (at $t\sim2.59$ Gyr) is not consistent with the observed properties. 

Among our models, the disruption phase (at $t\sim2.62$ Gyr) of the DYN arm is qualitatively the best at reproducing the observed trends. The correlation coefficients between velocity are both positive as observed, although the correlation is stronger for $V_{\rm pec}$-$d_{\rm Per}$ in the observational data (Figure~\ref{fig:CompModels}a). As shown in Figures~\ref{fig:CompModels}b and \ref{fig:CompModels}c, both $\langle U_{\rm pec}\rangle$ and $\langle V_{\rm pec}\rangle$ are also in good agreement with our observational results, and the trailing side shows higher values than the leading side. The {\rm observed} negative vertex deviation is also reproduced in the trailing side. On the other hand, the leading side shows less sensitivity of vertex deviation to the phase of the DYN arm, i.e. always positive, which is inconsistent with our observed trend (Figure~\ref{fig:CompModels}d). However, the measurement of vertex deviation in the leading side is less reliable. Hence, the disruption phase of the DYN arm shows qualitative agreement with the high-confidence results of our Cepheid data. Considering that our $N$-body simulations are still far from the real Milky Way because of lack of physical processes and lack of observational constraints, it is striking to find this level of agreement between the disruption phase of our simulated DYN arm and the observed trends found in our Cepheid data. We therefore conclude that the disruption phase of a DYN spiral arm like seen in our simulation, is the most likely scenario for the Perseus arm in the Milky Way. 

It is known that the age of Cepheids is well correlated with their pulsation period ($\log P$) \citep[e.g.][]{Bono+2005}. We color coded Cepheids by $\log P$ in Figure~\ref{fig:ObsCephMap}. We found no clear correlation between the age of Cepheids with the position with respect to the arm. The SDW scenario predicts a clear correlation between the age of stars and the distance from the arm \citep[e.g.][]{DobbsPringle2010}. Hence, this is also against the prediction from the SDW scenario, but more consistent with the DYN arm scenario \citep[e.g.][]{Grand+2012b}.

Interestingly, according to \citet{Reid+2014a}, the pitch angle of the Perseus arm ($9.4\pm1.4$~deg) is smaller than the Scutum arm ($19.8\pm2.6$~deg), which is the other major arm. $N$-body simulations of DYN arms predict that the pitch angle of spiral arms in the disruption phase would be smaller, because the arms are winding and disrupting \citep{Baba+2013,Grand+2013}. Therefore, if the Perseus arm does indeed have a small pitch angle, it is also consistent with the arm being in the disruption phase. We will further test the disruption phase scenario of the Perseus arm with the future Gaia data releases.

\acknowledgments
We are grateful to the referee for useful suggestions that helped improve this manuscript.
We also thank Nobuyuki Sakai for much useful advice on analysis of astrometric data. 
We thank Jo Bovy for making {\tt galpy} \citep{Bovy2015}, which is used for coordinate transformation, publicly available.
JB was supported by the Japan Society for the Promotion of Science (JSPS) 
Grant-in-Aid for Young Scientists (B) Grant Number 26800099.
DK acknowledges the support of the UK's Science \& Technology Facilities Council (STFC Grant ST/N000811/1). 
NM is grateful to Grant-in-Aid (KAKENHI, No.~26287028) from the Japan Society for the Promotion of Science (JSPS). 
JASH is supported by a Dunlap Fellowship at the Dunlap Institute for Astronomy \& Astrophysics, funded through an endowment established by the Dunlap family and the University of Toronto.
RG acknowledges support by the DFG Research Centre SFB-881 `The Milky Way System' through project A1.
This work has made use of data from the European Space Agency (ESA) mission 
Gaia (https://www.cosmos.esa.int/gaia), 
processed by the Gaia Data Processing and Analysis Consortium 
(DPAC, https://www.cosmos.esa.int/web/gaia/dpac/consortium). 
Funding for the DPAC has been provided by national institutions, 
in particular the institutions participating in the Gaia Multilateral Agreement. 
The simulations reported in this paper were carried out on facilities of 
Center for Computational Astrophysics (CfCA), National Astronomical Observatory of Japan.

\end{document}